\documentclass{article}
\usepackage{spconf,amsmath,epsfig}
\ninept

\title{Adaptive Reduced-Rank RLS Algorithms based on Joint Iterative
Optimization of Filters for Space-Time Interference Suppression}

\name{\fontsize{11}{11}\selectfont\upshape Rodrigo C. de Lamare
$\dagger$ and Raimundo Sampaio-Neto \vspace{-1.25em}
 $\ddagger$ %\thanks{Thanks to XYZ agency for funding.}
}
\address{\fontsize{10}{10}\selectfont\itshape  $\dagger$  Communications Research Group, University of York, United Kingdom \\
\fontsize{10}{10}\selectfont\itshape $\ddagger$  CETUC, Pontifical
Catholic University of Rio de Janeiro
(PUC-RIO), Brazil \\
\fontsize{9}{9}\selectfont\ttfamily\upshape E-mails:
rcdl500@ohm.york.ac.uk, raimundo@cetuc.puc-rio.br}

\begin{document}

\maketitle

\begin{abstract}
This paper presents novel adaptive reduced-rank filtering
algorithms based on joint iterative optimization of adaptive
filters. The novel scheme consists of a joint iterative
optimization of a bank of full-rank adaptive filters that
constitute the projection matrix and an adaptive reduced-rank
filter that operates at the output of the bank of filters. We
describe least squares (LS) expressions for the design of the
projection matrix and the reduced-rank filter and recursive least
squares (RLS) adaptive algorithms for its computationally
efficient implementation. Simulations for a space-time
interference suppression in a CDMA system application show that
the proposed scheme outperforms in convergence and tracking the
state-of-the-art
reduced-rank schemes at about the same complexity.\\
\end{abstract}


\section{Introduction}

In adaptive filtering \cite{diniz}, there is a huge number of
algorithms with different trade-offs between performance and
complexity. Among them, recursive least squares (RLS) algorithms
arise as the preferred choice with respect to convergence
performance. A challenging problem which remains unsolved by
conventional techniques is that when the number of elements in the
filter is large, the algorithm requires a large number of samples
to reach its steady-state behavior. In these situations, even RLS
algorithms require an amount of data proportional to $2M$
\cite{diniz} in stationary environments, where $M$ is the filter
length, to converge and this may lead to unacceptable performance.
Reduced-rank filtering \cite{scharf}-\cite{delamaresp} is a
powerful and effective technique in low sample support situations
and in problems with large filters. The advantages of reduced-rank
adaptive filters are their faster convergence speed and better
tracking performance than existing techniques when dealing with
large number of weights. Furthermore, in dynamic scenarios large
filters usually fail or provide poor performance in tracking
signals embedded in interference. Several reduced-rank methods and
systems have been proposed in the last several years, namely,
eigen-decomposition techniques \cite{bar-ness}-\cite{song&roy},
the multistage Wiener filter (MWF) \cite{gold&reed,goldstein} and
the auxiliary vector filtering (AVF) algorithm \cite{avf}. The
main problem with the best known techniques is their high
complexity and the fact that there is no joint optimization of the
mapping that carries out dimensionality reduction and the
reduced-rank filter.

In this work we propose an adaptive reduced-rank filtering scheme
based on combinations of adaptive filters with RLS algorithms. The
novel scheme consists of a joint iterative optimization of a bank
of full-rank adaptive filters which constitutes the projection
matrix and an adaptive reduced-rank filter that operates at the
output of the bank of full-rank filters. The essence of the
proposed approach is to change the role of adaptive filters. The
bank of adaptive filters is responsible for performing
dimensionality reduction, whereas the reduced-rank filter
effectively estimates the desired signal. Despite the large
dimensionality of the projection matrix and its associated slow
learning behavior, the proposed and existing \cite{goldstein,avf}
reduced-rank techniques enjoy in practice a very fast convergence.
The reason is that even an inaccurate or rough estimation of the
projection matrix is able to provide an appropriate dimensionality
reduction for the reduced-rank filter, whose behavior will govern
most of the performance of the overall scheme. We describe least
squares (LS) expressions for the design of the projection matrix
and the reduced-rank filter along with RLS adaptive algorithms for
its computationally efficient implementation. The performance of
the proposed scheme is assessed via simulations for a space-time
interference suppression application in DS-CDMA systems.

This work is organized as follows. Section $2$ states the
reduced-rank estimation problem. Section $3$ presents the novel
reduced-rank scheme, the joint iterative optimization and the LS
design of the filters. Section $4$ derives RLS algorithms for
implementing the proposed scheme. Section $5$ shows and discusses
the simulations, while Section $6$ gives the conclusions.

\section{Reduced-Rank Least Squares Parameter Estimation and Problem Statement}

The exponentially weighted LS estimator is the parameter vector
${\bf w}[i]= [w_1^{[i]}~ w_2^{[i]} ~ \ldots ~ w_M^{[i]}]^T$, which
is designed to minimize the following cost function
\begin{equation}
{\mathcal{C}}  = \sum_{l=1}^i \lambda^{i-l} | d[l] - {\bf
w}^{H}[i]{\bf r}[l]|^2
\end{equation}
where $d[l]$ is the desired signal, ${\bf r}[i]=[r_{0}^{[i]}~
\ldots ~r_{M-1}^{[i]}]^{T}$ is the input data, $(\cdot)^{T}$ and
$(\cdot)^{H}$ denote transpose and Hermitian transpose,
respectively, and $\lambda$ stands for the forgetting factor. The
set of parameters ${\bf w}[i]$ can be estimated via standard
stochastic gradient or LS estimation techniques \cite{diniz}.
However, the laws that govern the convergence behavior of these
estimation techniques imply that the convergence speed of these
algorithms is proportional to $M$, the number of elements in the
estimator. Thus, large $M$ implies slow convergence. A
reduced-rank algorithm attempts to circumvent this limitation in
terms of speed of convergence by reducing the number of adaptive
coefficients and extracting the most important features of the
processed data. This dimensionality reduction is accomplished by
projecting the received vectors onto a lower dimensional subspace.
Specifically, consider an $M \times D$ projection matrix ${\bf
S}_{D}[i]$ which carries out a dimensionality reduction on the
received data as given by
\begin{equation}
\bar{\bf r}[i] = {\bf S}_D^H[i] {\bf r}[i]
\end{equation}
where, in what follows, all $D$-dimensional quantities are denoted
with a "bar". The resulting projected received vector $\bar{\bf
r}[i]$ is the input to a tapped-delay line filter represented by
the $D \time 1$ vector $\bar{\bf w}[i]=[ \bar{w}_1^{[i]}
~\bar{w}_2^{[i]}~\ldots\bar{w}_D^{[i]}]^T$ for time interval $i$.
The estimator output corresponding to the $i$th time instant is
\begin{equation}
x[i] = \bar{\bf w}^{H}[i]\bar{\bf r}[i]
\end{equation}
If we consider the LS design in (1) with the reduced-rank
parameters we obtain
\begin{equation}
\bar{\bf w}[i] = \bar{\bf R}^{-1}[i]\bar{\bf p}[i]
\end{equation}
where $\bar{\bf R}[i] = \sum_{l=1}^i \lambda^{i-l} \bar{\bf
r}[l]\bar{\bf r}^{H}[l]={\bf S}_D^H[i]{\bf R}[i]{\bf S}_D[i]$ is
the reduced-rank covariance matrix, ${\bf R}[i] = \sum_{l=1}^i
\lambda^{i-l}{\bf r}[l]{\bf r}^{H}[l]$ is the full-rank covariance
matrix, $\bar{\bf p}[i]=\sum_{l=1}^i \lambda^{i-l} d^*[l]\bar{\bf
r}[l]={\bf S}_D^H[i]{\bf p}[i]$ is the cross-correlation vector of
the reduced-rank model and the vector ${\bf p}[i]=\sum_{l=1}^i
\lambda^{i-l} d^*[l]{\bf r}[l]]$ is the cross-correlation vector
of the full-rank model. The associated sum of error squares (SES)
for a rank $D$ estimator is expressed by
\begin{equation}
\begin{split}
{\rm SES} & = \sigma^2_d - \bar{\bf p}^H[i] \bar{\bf
R}^{-1}[i]\bar{\bf p}[i] \\ & = \sigma^2_d - {\bf p}^H[i]{\bf
S}_D[i] ({\bf S}_D^H[i]{\bf R}[i]{\bf S}_D[i])^{-1} {\bf
S}_D^H[i]{\bf p}[i]
\end{split}
\end{equation}
where $\sigma^2_d=\sum_{l=1}^i \lambda^{i-l} | d(l)|^2$.
%In the Appendix, we provide a necessary and sufficient condition
%for a projection ${\bf S}_D$ with dimension $M \times D$ not to
%modify the SES and discuss the existence of multiple solutions.
Based upon the problem statement above, the rationale for
reduced-rank schemes can be simply put as follows. How to
efficiently (or optimally) design a transformation matrix ${\bf
S}_D[i]$ with dimension $M \times D$ that projects the observed
data vector ${\bf r}[i]$ with dimension $M \times 1$ onto a
reduced-rank data vector $\bar{\bf r}[i]$ with dimension $D \times
1$? In the next section we present the proposed reduced-rank
approach.

\section{Proposed Reduced-Rank Scheme and Least Squares Design}

In this section we detail the principles of the proposed
reduced-rank scheme using a projection operator based on adaptive
filters and present a least squares (LS) design approach for the
estimators. The novel scheme, depicted in Fig. 1, employs a
projection matrix ${\bf S}_{D}[i]$ with dimension $M \times D$,
that is responsible for the dimensionality reduction, to process a
data vector $\bar{\bf r}[i]$ with dimension $M \times 1$ and map
it into a reduced-rank  data vector $\bar{\bf r}[i]$. The
reduced-rank filter $\bar{\bf w}[i]$ with dimension $D \times 1$
processes the reduced-rank data vector $\bar{\bf r}[i]$ in order
to yield a scalar estimate $x[i]$. The projection matrix ${\bf
S}_{D}[i]$ and the reduced-rank filter $\bar{\bf w}[i]$ are
jointly optimized in the proposed scheme according to the LS
criterion.

\begin{figure}[htb]
       \centering  % figura centralizada
       \hspace*{-2.75em}
        \vspace*{-2em}
      {\includegraphics[width=11cm, height=4.0cm]{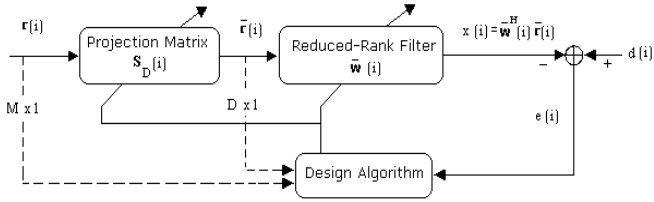}}
       \vspace*{-1em}
       \caption{\small Proposed Reduced-Rank Scheme.}
\end{figure}

Specifically, the projection matrix is structured as a bank of $D$
full-rank filters ${\bf s}_d[i]=\big[s_{1,d}^{[i]} ~
s_{2,d}^{[i]}~ \ldots~s_{M,d}^{[i]} \big]^T$, $d = 1,~\ldots,~D$,
with dimensions $M \times 1$ as given by ${\bf S}_{D}[i] =
\big[~{\bf s}_1^{[i]} ~| ~{\bf s}_2^{[i]}~| ~\ldots~|{\bf
s}_D^{[i]}~\big]$. The output estimate $x[i]$ of the reduced-rank
scheme as a function of the received data ${\bf r}[i]$, the
projection matrix ${\bf S}_D[i]$ and the reduced-rank filter
$\bar{\bf w}[i]$ is
\begin{equation}
\begin{split}
x[i] & =  \bar{\bf w}^H[i] {\bf S}_D^H[i] {\bf r}[i] = \bar{\bf
w}^H[i] \bar{\bf r}[i]
\end{split}
\end{equation}
Note that for $D=1$, the novel scheme becomes a conventional
full-rank filtering scheme with an addition weight parameter $w_D$
that provides a gain. For $D>1$, the signal processing tasks are
changed and the full-rank filters compute a subspace projection
and the reduced-rank filter estimates the desired signal.

We describe LS expressions for the design of the projection matrix
and the reduced-rank filter along with RLS adaptive algorithms for
its computationally efficient implementation. Let us consider the
exponentially-weighted LS expressions for the filters ${\bf
S}_D[i]$ and $\bar{\bf w}[i]$ can be computed via the cost
function given by
\begin{equation}
\begin{split}
{\mathcal{C}} & = \sum_{l=1}^i \lambda^{i-l} | d[l] - \bar{\bf
w}^{H}[i]{\bf S}_D^H[i]{\bf
r}(l)|^2 \big] %\\
%& \quad +  \sum_{j=1}^i \lambda^{i-j} | d(j) - \bar{\bf
%w}^{H}(j){\bf S}_D^H[i]{\bf r}(j)|^2
\end{split}
\end{equation}
By minimizing (7) with respect to $\bar{\bf w}[i]$, the
reduced-rank filter weight vector becomes
\begin{equation}
\bar{\bf w}[i] = \bar{\bf R}^{-1}[i] \bar{\bf p}[i]
\end{equation}
where $\bar{\bf p}[i] = {\bf S}_D^H[i] \sum_{l=1}^i
\lambda^{i-l}d^{*}[l]{\bf r}[l] = \sum_{l=1}^i
\lambda^{i-l}d^{*}[l]\bar{\bf r}[l]]$, $\bar{\bf R}[i] = {\bf
S}_D^H[i]\sum_{l=1}^i \lambda^{i-l}{\bf r}[l]{\bf r}^H[l] {\bf
S}_D[i] $. By minimizing (7) with respect to ${\bf S}_D[i]$ we
obtain
\begin{equation} {\bf S}_D[i] = {\bf R}^{-1}[i] {\bf P}_D[i] {\bf
R}_{w}^{-1}[i]
\end{equation}
where ${\bf P}_D[i] = \sum_{l=1}^i \lambda^{i-l}d^{*}[l]{\bf
r}[l]{\bf w}^H[i]$, the covariance matrix is ${\bf R}[i] =
\sum_{l=1}^i \lambda^{i-l}{\bf r}[l]{\bf r}^{H}[l]$ and ${\bf
R}_w[i] = \sum_{l=1}^i \lambda^{i-l}{\bf w}[l]{\bf w}^{H}[l]$. The
associated SES is
\begin{equation}
{\rm SES} = \sigma^{2}_{d} - \bar{\bf p}^{H}[i] \bar{\bf
R}^{-1}[i] \bar{\bf p}[i]
\end{equation}
where $\sigma^{2}_{d}=\sum_{l=1}^i \lambda^{i-l}|d[l]|^{2}$. Note
that the expressions in (8) and (9) are not closed-form solutions
for $\bar{\bf w}[i]$ and ${\bf S}_D[i]$ since (8) is a function of
${\bf S}_D[i]$ and (9) depends on $\bar{\bf w}[i]$ and thus they
have to be iterated with an initial guess to obtain a solution.
The key strategy lies in the joint optimization of the filters.
The rank $D$ must be set by the designer to ensure appropriate
performance. The expressions in (8) and (9) require the inversion
of matrices. In order to reduce the complexity, we employ the
matrix inversion lemma and derive RLS algorithms in the next
section. The rank $D$ must be set by the designer to ensure
appropriate performance and the reader is referred to \cite{qian}
for rank selection methods. In the next section, we seek iterative
solutions via adaptive algorithms.

\section{Proposed RLS Algorithms}

In this section we propose RLS algorithms for efficiently
implementing the LS design of the previous section. Firstly, let
us consider the expression in (8) with its associated quantities,
i.e. the matrix $\bar{\bf R}[i] = \sum_{l=1}^i
\lambda^{i-l}\bar{\bf r}[l] \bar{\bf r}^{H}[l]$ and the vector
$\bar{\bf p}[i] = \sum_{l=1}^i \lambda^{i-l}d^{*}[l]\bar{\bf
r}[l]$, define $\boldsymbol{\bar{\Phi}}[i] = {\bf R}^{-1}[i]$ and
rewrite $\bar{\bf p}[i]$ as $\bar{\bf p}[i] = \lambda \bar{\bf
p}[i-1] + d^*[i] \bar{\bf r}[i]$. We can write (8) in an
alternative form as follows
\begin{equation}
\begin{split}
\bar{\bf w}[i] & = \boldsymbol{\bar{\Phi}}[i] \bar{\bf p}[i] =
\lambda \boldsymbol{\bar{\Phi}}[i] \bar{\bf p}[i-1] +
\boldsymbol{\bar{\Phi}}[i]\bar{\bf r}[i] d^*[i] \\
& = \boldsymbol{\bar{\Phi}}[i-1] \bar{\bf p}[i-1] - \bar{\bf k}[i]
\bar{\bf r}^H[i] \boldsymbol{\bar{\Phi}}[i-1] \bar{\bf p}[i-1] +
\boldsymbol{\bar{\Phi}}[i] \bar{\bf r}[i] d^*[i] \\
& = \bar{\bf w}[i-1] - \bar{\bf k}[i] \bar{\bf r}^H[i] \bar{\bf
w}[i-1] + \bar{\bf k}[i] d^*[i] \\
& = \bar{\bf w}[i-1] + \bar{\bf k}[i] \big[ d^*[i] - \bar{\bf
r}^H[i] \bar{\bf w}[i-1] \big]
\end{split}
\end{equation}
By defining $\xi[i] = d[i] - \bar{\bf w}^H[i-1]\bar{\bf r}^H[i] $
we arrive at the proposed RLS algorithm for estimating $\bar{\bf
w}[i]$
\begin{equation}
\bar{\bf w}[i] = \bar{\bf w}[i-1] + \bar{\bf k}[i] \xi^*[i]
\end{equation}
where the so-called Kalman gain vector is given by
\begin{equation}
\bar{\bf k}[i] = \frac{\lambda^{-1} \boldsymbol{\bar{\Phi}}[i-1]
\bar{\bf r}[i] }{1 + \lambda^{-1} \bar{\bf r}^H[i]
\boldsymbol{\bar{\Phi}}[i-1]\bar{\bf r}[i]}
\end{equation}
and the update for the matrix inverse $\boldsymbol{\bar{\Phi}}[i]$
employs the matrix inversion lemma \cite{diniz}
\begin{equation}
\boldsymbol{\bar{\Phi}}[i] = \lambda^{-1}
\boldsymbol{\bar{\Phi}}[i-1] - \lambda^{-1} \bar{\bf k}[i]
\bar{\bf r}^H[i] \boldsymbol{\bar{\Phi}}[i-1]
\end{equation}
Note that the proposed RLS algorithm given in (12)-(14) is similar
to the conventional RLS algorithm \cite{diniz}, except that it
works in a reduced-rank model with a $D \times 1$ input $\bar{\bf
r} [i] = {\bf S}_D^H [i] {\bf r}[i]$, where the $M \times D$
matrix ${\bf S}_D$ is the projection matrix responsible for
dimensionality reduction. Now let us present the second part of
the proposed RLS algorithms, in which we detail the design of
${\bf S}_D[i]$. Let us define ${\bf P}[i] ={\bf R}^{-1}[i]$, ${\bf
Q}_{\bar{\bf w}}[i-1]={\bf P}^{-1}_{\bar{\bf w}}[i]$, ${\bf P}_D
[i] = \lambda {\bf P}_{D}[i-1] + d^*[i] {\bf r}[i] {\bf w}^H [i]$
and rewrite the expression in (9) as follows
\begin{equation}
\begin{split}
{\bf S}_D [i] & = \hat{\bf R}[i] {\bf P}_D [i] {\bf P}_{\bf
w}[i-1] = {\bf P}[i] {\bf P}_D[i] {\bf Q}_{\bar{\bf w}}[i-1] \\
& = \lambda {\bf P}[i] {\bf P}_D [i-1] {\bf Q}_{\bar{\bf w}}[i-1]
+ d^*[i] {\bf P}[i] {\bf r}[i] \bar{\bf w}^H[i] {\bf Q}_{\bar{\bf
w}}[i] \\ & = {\bf S}_D[i-1] - {\bf k}[i] {\bf P}[i-1] {\bf
P}_D[i-1] {\bf Q}_{\bar{\bf w}}[i] \\ & \quad + d^*[i]{\bf P}[i]
{\bf r}[i] \bar{\bf w}^H[i] {\bf Q}_{\bar{\bf w}}[i] \\
& = {\bf S}_D[i-1] - {\bf k}[i] {\bf P}[i-1] {\bf P}_D[i-1] {\bf
Q}_{\bar{\bf w}}[i] \\ & \quad + d^*[i]{\bf k}[i]
 \bar{\bf w}^H[i] {\bf Q}_{\bar{\bf w}}[i] \\
\end{split}
\end{equation}
By defining the vector ${\bf t}[i] = {\bf Q}_{\bar{\bf w}}[i]
\bar{\bf w}[i]$ and using the fact that $\bar{\bf r}^H[i-1] = {\bf
r}^H[i-1]{\bf S}_D[i-1]$ we arrive at
\begin{equation}
{\bf S}_D[i] = {\bf S}_D[i-1] + {\bf k}[i] \big( d^*[i] {\bf
t}^H[i] - \bar{\bf
 r}^H[i] \big)
\end{equation}
where the Kalman gain vector for the estimation of ${\bf S}_D[i]$
is
\begin{equation}
{\bf k}[i] = \frac{\lambda^{-1} {\bf P}[i-1] {\bf r}[i] }{1 +
\lambda^{-1} {\bf r}^H[i] {\bf P}[i-1]{\bf r}[i]}
\end{equation}
and the update for the matrix ${\bf P}[i]$ employs the matrix
inversion lemma \cite{diniz}
\begin{equation}
{\bf P}[i] = \lambda^{-1} {\bf P}[i-1] - \lambda^{-1} {\bf k}[i]
{\bf r}^H[i] {\bf P}[i-1]
\end{equation}
the vector ${\bf t}[i]$ is updated as follows
\begin{equation}
{\bf t}[i] = \frac{\lambda^{-1} {\bf Q}_{\bar{\bf w}}[i-1]
\bar{\bf w}[i-1] }{1 + \lambda^{-1} \bar{\bf w}^H[i-1] {\bf
Q}_{\bar{\bf w}}[i-1]\bar{\bf w}[i-1]}
\end{equation}
and the matrix inversion lemma is used to update ${\bf
Q}_{\bar{\bf w}}[i]$ as described by
\begin{equation}
{\bf Q}_{\bar{\bf w}}[i] = \lambda^{-1} {\bf Q}_{\bar{\bf w}}[i-1]
- \lambda^{-1} {\bf t}[i] \bar{\bf w}^H[i-1]
\end{equation}
The equations (16)-(20) constitute the second part of the proposed
RLS algorithms and are responsible for estimating the projection
matrix ${\bf S}_D[i]$. The computational complexity of the
proposed RLS algorithms is $O(D^2)$ for the estimation of
$\bar{\bf w}[i]$ and $O(M^2)$ for the estimation of ${\bf
S}_D[i]$. Because $D << M$, as will be explained in the next
section, the overall complexity is in the same order of the
conventional full-rank RLS algorithm ($O(M^2)$) \cite{diniz}.

\section{Simulations}

The performance of the proposed scheme is assessed via simulations
for space-time CDMA interference suppression. We consider the
uplink of DS-CDMA system with symbol interval $T$, chip period
$T_c$, spreading gain $N=T/Tc$, $K$ users and equipped with $J$
elements in a uniform antenna array. The spacing between the
antenna elements is $d=\lambda_c/2$, where $\lambda_c$ is carrier
wavelength. Assuming that the channel is constant during each
symbol and the base station receiver is synchronized with the main
path, the received signal after filtering by a chip-pulse matched
filter and sampled at chip rate yields the $JM\times 1$ received
vector
\begin{equation}
\begin{split}
{\bf r}[i] & =  \sum_{k=1}^{K} A_{k}b_{k}[i-1] \bar{\bf
p}_{k}[i-1] + A_{k}b_{k}[i] {\bf p}_{k}[i] \\ & \quad +
A_{k}b_{k}[i+1] \tilde{\bf p}_{k}[i+1] + {\bf
 n}[i],
\end{split}
\end{equation}
where $M=N+L_p-1$, the complex Gaussian noise vector is ${\bf
n}[i] = [n_{1}[i] ~\ldots~n_{JM}[i]]^{T}$ with $E[{\bf n}[i]{\bf
n}^{H}[i]] = \sigma^{2}{\bf I}$, $(\cdot)^{T}$ and $(\cdot)^{H}$
denote transpose and Hermitian transpose, respectively, and
$E[\cdot]$ stands for expected value. The spatial signatures are
$\bar{\bf p}_{k}[i-1] = \boldsymbol{\bar{\mathcal F}}_{k}
\boldsymbol{{\mathcal H}}_{k}[i-1]$, ${\bf p}_{k}[i] =
\boldsymbol{{\mathcal F}}_{k} \boldsymbol{{\mathcal H}}_{k}[i]$
and $\tilde{\bf p}_{k}[i] = \boldsymbol{\tilde{\mathcal F}}_{k}
\boldsymbol{{\mathcal H}}_{k}[i+1]$, where
$\boldsymbol{\bar{\mathcal F}}_{k}$, $\boldsymbol{{\mathcal
F}}_{k}$ and $\boldsymbol{\tilde{\mathcal F}}_{k}$ are block
diagonal matrices with one-chip shifted versions of segments of
the signature sequence ${\bf s}_{k} = [a_{k}(1) \ldots
a_{k}(N)]^{T}$ of user $k$. The $JL_p \times 1$ space-time channel
vector is given by $\boldsymbol{{\mathcal H}_{k}}[i] = \big[{\bf
h}_{k,0}^{T}[i] |~{\bf h}_{k,1}^{T}[i]|  ~\ldots ~|{\bf
h}_{k,J-1}^{T}[i]\big]^{T}$ with ${\bf h}_{k,l}[i] =
[h_{k,0}^{(l)}[i] \ldots h_{k,L-1}^{(l)}[i]]^{T}$ being the
channel of user $k$ at antenna element $l$ with their associated
DoAs $\phi_{k,m}$.

For all simulations, we use the initial values $\bar{\bf w}[0]=
[1~0~\ldots~0]^T$ and ${\bf S}_D[0]=[{\bf I}_D ~ {\bf
0}_{D,JM-D}]^T$, assume $L=9$ as an upper bound, use $3$-path
channels with relative powers given by $0$, $-3$ and $-6$ dB,
where in each run the spacing between paths is obtained from a
discrete uniform random variable between $1$ and $2$ chips and
average the experiments over $200$ runs. The DOAs of the
interferers are uniformly distributed in $(0,2\pi/3)$. The system
has a power distribution among the users for each run that follows
a log-normal distribution with associated standard deviation equal
to $1.5$ dB. We compare the proposed scheme with the Full-rank
\cite{diniz}, the MWF \cite{goldstein} and the AVF \cite{avf}
techniques for the design of linear receivers, where the
reduced-rank filter $\bar{\bf w}[i]$ with $D$ coefficients
provides an estimate of the desired symbol for the desired used
(user $1$ in all experiments) using the bit error rate (BER)
\cite{goldstein}. We consider the BER performance versus the rank
$D$ with optimized parameters (forgetting factors $\lambda=0.998$)
for all schemes. The results in Fig. 2 indicate that the best rank
for the proposed scheme is $D=4$ (which will be used in the
remaining experiments) and it is very close to the optimal MMSE.
Studies with systems with different processing gains show that $D$
is invariant to the system size, which brings considerable
computational savings.

\begin{figure}[!htb]
\begin{center}
\def\epsfsize#1#2{1\columnwidth}
\epsfbox{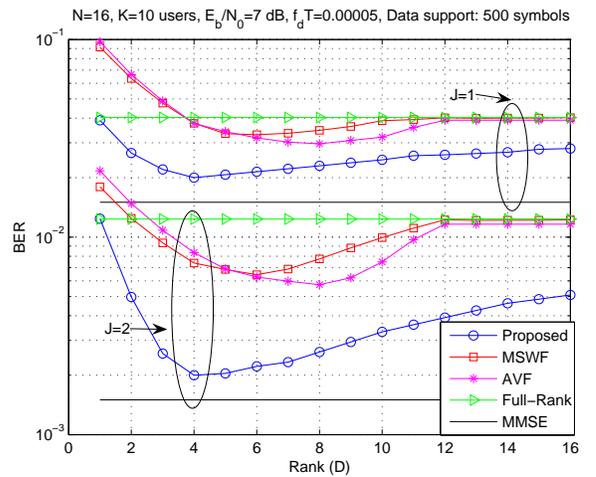} \caption{\small BER performance versus rank (D).}
\end{center}
\end{figure}

We compare the proposed scheme with the Full-rank \cite{diniz},
the MWF \cite{goldstein} and the AVF \cite{avf} techniques for the
design of linear receivers, where the reduced-rank filter
$\bar{\bf w}[i]$ with $D$ coefficients provides an estimate of the
desired symbol for the desired used (user $1$ in all experiments)
using the signal-to-interference-plus-noise ratio (SINR)
\cite{goldstein}. We consider the BER performance versus the rank
$D$ with optimized parameters (forgetting factors $\lambda$) for
all schemes. The results in Fig. 2 indicate that the best rank for
the proposed scheme is $D=4$ (which will be used in the remaining
experiments) and it is very close to the optimal MMSE. Studies
with systems with different processing gains show that $D$ is
relatively invariant to the system size, which brings considerable
computational savings to the proposed scheme and allows a very
fast convergence performance. In practice, the rank $D$ can be
adapted in order to obtain fast convergence and ensure good steady
state performance and tracking after convergence.

\begin{figure}[!htb]
\begin{center}
\def\epsfsize#1#2{1\columnwidth}
\epsfbox{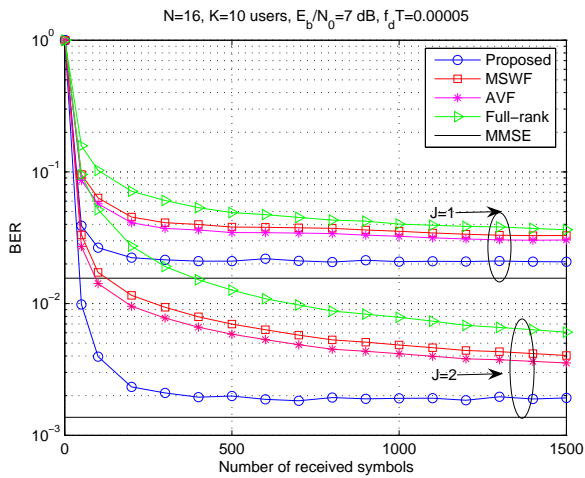} \caption{\small BER performance versus number of
received symbols.}
\end{center}
\end{figure}

The BER convergence performance in a mobile communications
situation is shown in Fig. 3. The channel coefficients are
obtained with Clarke´s model \cite{rappa} and the adaptive filters
of all methods are trained with $200$ symbols and then switch to
decision-directed mode. The results show that the proposed scheme
has a much better performance than the existing approaches and is
able to adequately track the desired signal. A complete
convergence analysis of the proposed scheme, including tracking
and steady-state performance, conditions and proofs are not
included here due to lack of space and are intended for a future
paper.

\section{Conclusions}

We proposed a novel reduced-rank scheme based on joint iterative
optimization of filters with an implementation using RLS
algorithms. In the proposed scheme, the full-rank adaptive filters
are responsible for estimating the subspace projection rather than
the desired signal, which is estimated by a small reduced-rank
filter. The results for space-time interference suppression in a
DS-CDMA system show a performance significantly better than
existing schemes and close to the optimal MMSE in a dynamic and
hostile environment.

\end{document}